\documentclass[conference]{IEEEtran}
\IEEEoverridecommandlockouts
\usepackage{cite}
\usepackage{amsmath,amssymb,amsfonts}
\usepackage{algorithmic}
\usepackage{graphicx}
\usepackage{textcomp}
\usepackage{xcolor}
\def\BibTeX{{\rm B\kern-.05em{\sc i\kern-.025em b}\kern-.08em
    T\kern-.1667em\lower.7ex\hbox{E}\kern-.125emX}}
\begin{document}

\title{Monitoring My Dehydration: A Non-Invasive Dehydration Alert System Using Electrodermal Activity}

\author{\IEEEauthorblockN{Nandan Kulkarni, Christopher Compton, Jooseppi Luna, Mohammad Arif Ul Alam}
\IEEEauthorblockA{
\textit{Department of Computer Science}\\
University of Massachusetts Lowell, USA \\
nandan\_kulkarni@student.uml.edu, christopher\_compton@student.uml.edu, jooseppi\_luna@student.uml.edu,  \\ mohammadariful\_alam@uml.edu}

}

\maketitle

\begin{abstract}
Staying hydrated and drinking fluids is extremely crucial to stay healthy and maintaining even basic bodily functions. Studies have shown that dehydration leads to loss of productivity, cognitive impairment and mood in both men and women. However, there are no such an existing tool that can monitor dehydration continuously and provide alert to users before it affects on their health. In this paper, we propose to utilize wearable Electrodermal Activity (EDA) sensors in conjunction with signal processing and machine learning techniques to develop first time ever a dehydration self-monitoring tool, \emph{Monitoring My Dehydration} (MMD), that can instantly detect the hydration level of human skin. Moreover, we develop an Android application over Bluetooth to connect with wearable EDA sensor integrated wristband to track hydration levels of the users real-time and instantly alert to the users when the hydration level goes beyond the danger level. To validate our developed tool's performance, we recruit 5 users, carefully designed the water intake routines to annotate the dehydration ground truth and trained state-of-art machine learning models to predict instant hydration level i.e., well-hydrated, hydrated, dehydrated and very dehydrated. Our system provides an accuracy of 84.5\% in estimating dehydration level with an sensitivity of 87.5\% and a specificity of 90.3\% which provides us confidence of moving forward with our method for larger longitudinal study.
\end{abstract}


\section{Introduction}
Hydration level is a strong indicator of health that can help improve medical implications on potential health hazards and it is extremely important to track the hydration level (HL) of human body, specifically for children, the elderly and patients with underlying medical conditions such as diabetes. Despite increased risks of disability, mortality and hospital admissions associated with water-loss, dehydration is often unnoticed due to lack of immediate symptoms and instant measurement that necessitates the dehydration measurement tool significantly.

This paper presents MMD, a wearable sensor technology in conjunction with machine learning and signal processing methods to continuously monitor hydration level of users in a non-invasive way. MMD uses an Electrodermal Activity (EDA) sensor integrated wearable wristband on human subjects, which, studies have shown, can be linked to moisture levels in the skin. Dehydration has been shown to affect a majority of adults \cite{1}, affecting concentration and mood \cite{2,3}, and deteriorating job performance and happiness \cite{2,3}. However, current techniques for measuring hydration are either expensive, inconvenient, inaccurate or invasive. One of the more expensive `gold standard' techniques involves having the subject ingest an isotope in a known amount, and then calculating the concentration of the isotope in a bodily fluid to determine the relative amount of water in the body. Less complicated (but less precise) techniques involve weighing the subject or taking urine samples \cite{4}. 

There has been limited research on using mobile and wearable sensors to detect dehydration; however, with continued improvement and introduction of new sensors, future prospects for this are promising. One new device that has been recently introduced is the Empatica E4 wristband \cite{12}, a device that connects to a smartphone via Bluetooth and detects a comprehensive set of physiological indicators such as EDA, heart rate, movement, temperature, and time \cite{5}.

This paper combines the EDA sensor of the E4 device with common machine-learning models and basic Android application frameworks to build a comprehensive application to detect user hydration levels and alert the user that he should be drinking water. The basic design of the system is presented as a flowchart (Fig. \ref{fig:overview}) with an explanation of all major parts of its implementation, including the communication method with the Empatica device and the notifications mechanism. Then data collection methods are described, along with difficulties encountered with both the lack of available participants and technical issues from wearing the device for long periods of time. Once this raw data is acquired, it has to be pre-processed; the pre-processing is described, and then details are given on all the different techniques used to train models. Finally, the results are presented along with conclusions and future work.

\section{Related Works}
Related works of two kinds are considered: studies and articles on the prevalence of dehydration and its effect on productivity are briefly discussed, followed by a discussion of works on the use of sensors to detect dehydration. We used Empatica E4 and EDA sensor before for multi-label activity recognition \cite{alam1}, multiple person's activity recognition \cite{alam2} and functional, behavioral and cognition health assessment of older adults \cite{alam3}.
\subsection{Dehydration}
There is research that shows that dehydration is prevalent among Americans; for example, a 1998 study found that 75\% of Americans most likely experience a net fluid loss, therefore leading to dehydration \cite{1}. Furthermore, this dehydration leads to loss of productivity, impairing cognitive function and mood in both men and women \cite{2,3}. This happens not just to athletes or construction workers, but also to office workers and other tamer professions. Therefore, it can be argued that dehydration is a widespread phenomenon that affects the health, happiness, and productivity of Americans, and that most people are apparently unaware of it.
\subsection{Detecting Dehydration}
Dehydration detection using mobile sensors is a relatively new concept. In one study from 2019 \cite{5}, users in good health were recruited to initially perform a cognitive task known as the Stroop Task while being fully hydrated. During the course of the task, EDA and Pressure Relief Value readings were collected using a wearable sensor. The participants were then instructed to not consume liquids or water-heavy foods for the next 24 hours. Upon their return, they were instructed to perform the same task while wearing the sensors. Finally, the participants rehydrated and performed the same task once more. The authors then used various machine learning methods, such as logistic regression, support vector machines, decision trees, and K-Nearest Neighbor (KNN) classifiers to model the data and predict the dehydration of the user. Additionally, a variety of physiological measures were taken of the subjects to confirm that they were, indeed, mildly dehydrated. The authors mention at the end of the paper that one of the shortcomings of their study is that the hydration was only being sensed in a very controlled environment, and that it would be valuable future work to assess the accuracy of similar methods at determining hydration levels in less-controlled environments. Another somewhat-similar paper is concerned with extracting features from EDA data using a variety of methods and developed a new algorithm for the fast and efficient interpretation of EDA data into EDA \cite{6}.

\begin{figure}
  \centering
  \includegraphics[width=\linewidth]{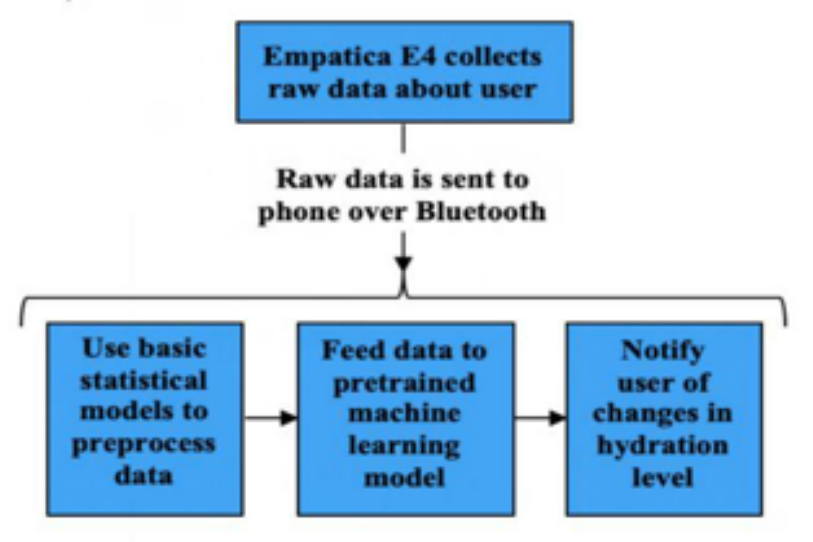}
  \caption{MMD Hydration Alert Android System Overview}
  \label{fig:overview}
\end{figure}

\section{Methodology}
In this section, the methodology of the project is laid out. It starts with a description of the overall layout of the application design and frameworks followed by a discussion of the data collection and preprocessing methods. Finally, the machine learning models created are discussed. The Empatica E4 sensor described in the introduction is used for all data collection.
\subsection{Overall Framework}
The overall project layout is shown in Fig. \ref{fig:overview}. The flow in Fig. \ref{fig:overview} starts with the Empatica E4 device, which collects raw data about the user and does some basic preprocessing (for example, calculating Skin Conductance from EDA signals, non-negative sparse deconvolution to extract components of EDA signal). It then sends this data to the user's smartphone over Bluetooth. The Android application on the user's phone then preprocesses the data by using basic statistical methods to help remove noise from the data. Once this is done, the data is fed to a pre-trained machine learning model using the Waikato Environment for Knowledge Analysis (WEKA) \cite{8} Java library. Then, if the machine learning model predicts a change in hydration level, it will trigger a method that sends the user a notification to alert them of their changed hydration level.

\begin{figure}
  \centering
  \includegraphics[width=\linewidth]{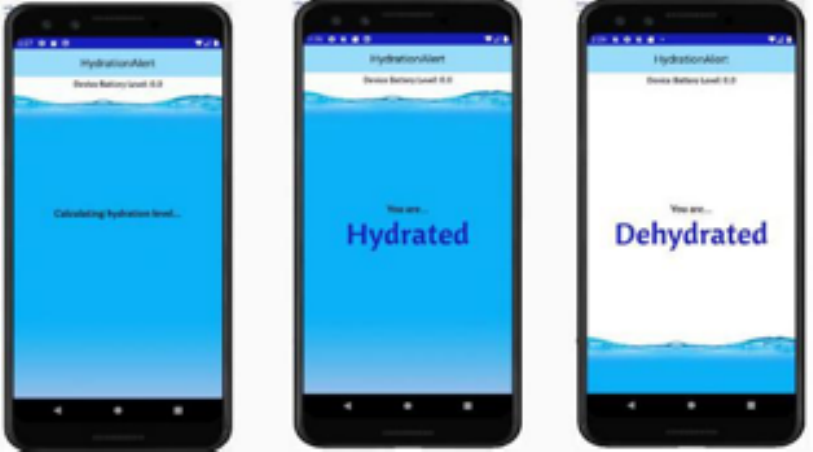}
  \caption{Application's user interface}
  \label{fig:alert_interface}
\end{figure}
\subsection{Electrodermal Activity Feature Extraction}
Electrodermal activity also known as skin conductance measurement over time includes two components. (i) Skin conductance Base Level (SBL), which changes slowly over time (tonic changes) and indicates the general activation of the sympathetic nervous system, (ii) Skin Conductance Responses (SCRs), changes that last for shorter periods (phasic changes). SCRs indicate the activation of the somatic nervous system (SNS) but also reflect responses to events that are new, unexpected, relevant, and/or aversive. Using EDA data to measure arousal in a continuous stimulus setting requires three steps in data processing and analysis. First step is pre-processing which involves data cleaning, filtering, downsampling, cutting, smoothing, artifact correction and decomposition of the signal into its tonic and phasic components. The SBL is typically approximated by frequency filtering, statistical modeling or simple linear interpolation between the skin conductance measures that are not overlaid by responses. The second step is parameterization, which involves deciding which parameter of the EDA data to measure/calculate. For a phasic parameter, this process includes massive abstraction of the phasic signal component, for example, counting responses. The third step is the correlation of the extracted data with the stimulus. We used LedaLab \cite{40} toolbox for EDA data preprocessing and extracting features. We employed butterworth low-pass filter, hanning smoothing with window size 4 and manual movement artifact correction. We decomposed EDA data into its tonic and phasic components using Continuous Decomposition Analysis (CDA) and Discrete Decomposition Analysis (DDA) as discussed below.

\begin{figure}
  \centering
  \includegraphics[width=\linewidth]{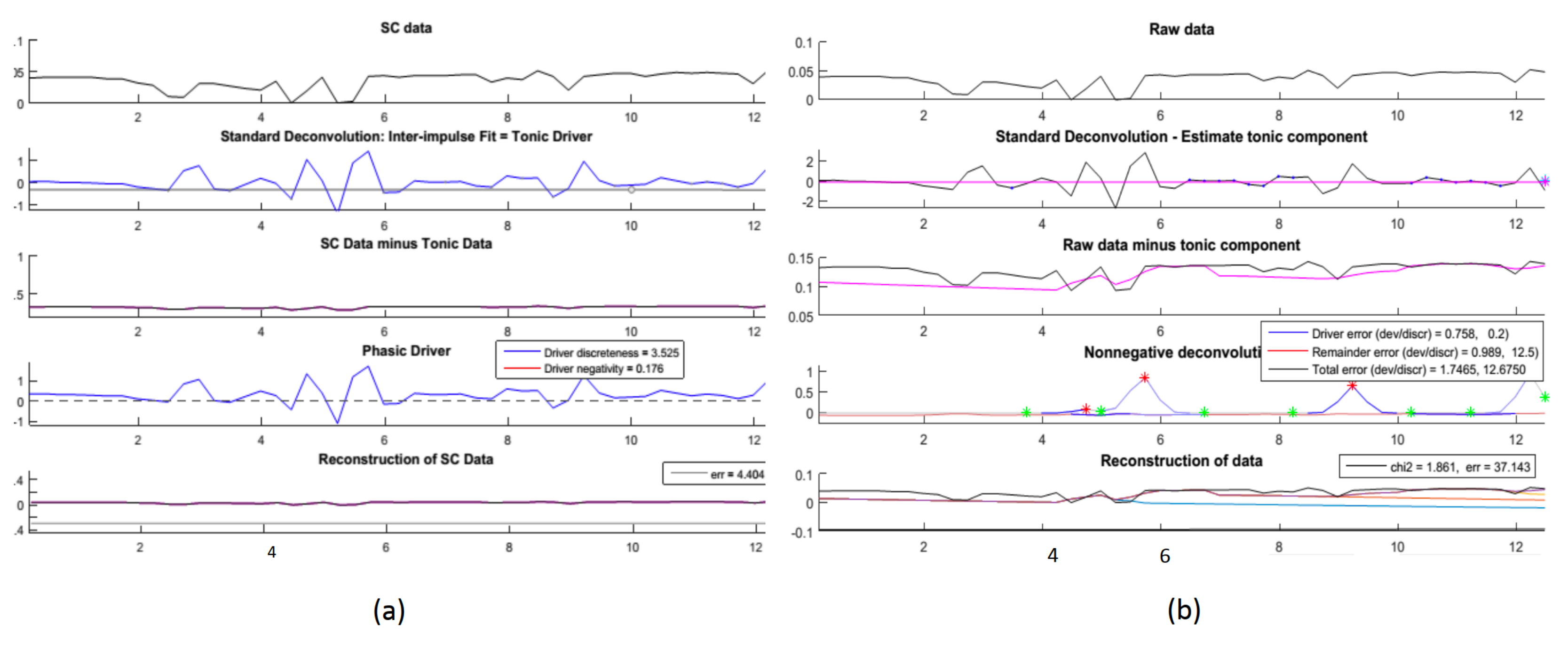}
  \caption{(a) CDA decomposition steps (b) DDA decomposition steps}
  \label{fig:decomposition}
\end{figure}

\subsubsection{Continuous Decomposition Analysis (CDA)}
This method helps extract the phasic (driver) information underlying the EDA signal, and aims at retrieving the signal characteristics of the underlying sudomotor nerve activity (SNA). EDA data is deconvolved by the general response shape which results in a large increase of temporal precision and then data is being decomposed into continuous phasic and tonic components \cite{38}. This helps compute the several standard features of phasic EDA. We tracked the related events as our pre-labeled activities and extracted 7 time-domain features from CDA. We used standard deviation, mean and variances on these features over the activity window. Fig. \ref{fig:decomposition}(a) shows different decomposition steps of CDA method of a single activity episode.
\subsubsection{Discrete Decomposition Analysis (DDA)}
This method decomposes EDA data into distinct phasic components and a tonic component by means of Nonnegative Deconvolution. The method helps capture and explore all intra-individual deviations of the general response shape and compute a detailed full model of all components in the entire data set \cite{39}. This method is particularly suited for physiological models of the SCR. We extracted 5 features from DDA for each activity window and extracted statistical mean, variance and standard deviation on these over the activity session. Fig. \ref{fig:decomposition}(b) shows different decomposition steps of DDA method of a single activity episode. The decomposition results in the extraction of distinct response components and thus allows for an unbiased quantification of Skin Conductance Response (SCR) characteristics (e.g., SCR amplitude). We extracted total 12 × 3 = 36 features from decomposition that are subject to be proportional to skin conductance or skin moisture level \cite{38,39}. We use state-of-art machine learning models to train the hydration level using the 36 features over an activity window.

\subsection{User Interface and Notifications}
The `HydrationAlert' application is relatively simple and is composed of the `MainActivity', a custom defined Service, and various other custom classes and enumerations. The main activity has three important functions. First, it is responsible for binding the Empatica E4 device service to the Android application, and handling any connection issues. Second, the main activity manages the UI for the device, displaying the relevant information in a clear, easy-to-understand interface. Third, it has the functions for creating and sending notifications; whenever the user's predicted hydration level changes, a callback function is triggered that both modifies the UI and also sends the user a notification. Whether the application may be active in the foreground or the background, connection always needs to be enabled until the application has been terminated.

Fig. \ref{fig:alert_interface} demonstrates a typical workflow for a user. We can see the interface the user will see when the user connects the device to the phone. Once connected, the phone initially calculates the hydration level and then displays it along with a visual cue of a high or low water level. The Empatica framework used for connecting to and receiving data from the device is implemented as a custom Android Service, implementing their public application programming interfaces. This interface allows the Android service to run in the background, even after the application is suspended for continuous monitoring of the user's EDA levels. A custom interface is defined within the Service as well to allow for the creation of delegate callback methods used by the main activity. The above-mentioned WEKA Java library is accessed as a Java class, which is subsequently instantiated by the Empatica Service. The Empatica Service processes the Electrodermal activity readings from the EDA sensor, passes them to the WEKA object, reads the results of the machine learning model, checks for a change in activity state, and triggers the callback function if there is a change in hydration level.

\begin{figure}
  \centering
  \includegraphics[width=\linewidth]{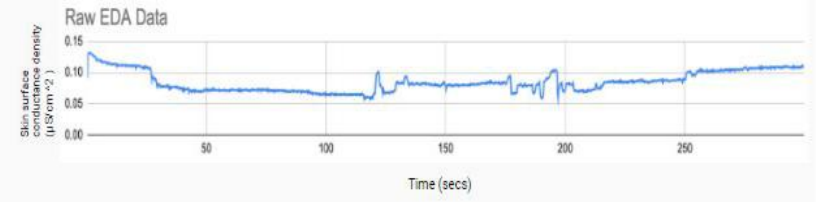}
  \caption{Raw EDA signal}
  \label{fig:raw_signal}
\end{figure}

\begin{table*}[h!]
  \begin{center}
    \caption{Comparisons of Classification Performances for multiple classifiers}
    \label{tab:classification}
    \begin{tabular}{ |c| c| c| c| c| c|}
\hline
  & Decision Tree & Random Forest & Naive Bayes & Multilayer perceptron & BayesNet\\ 
 \hline
 Accuracy & 84.5 $\pm0.1$& 83.2$\pm8.3$ & 70.3$\pm1.2$ & 80.2$\pm0.2$ & 69.2$\pm2.3$ \\ 
 \hline
 Sensitivity & 87.5$\pm0.1$ & 80.3$\pm1.3$ & 75.5$\pm1.8$ & 82.8$\pm3.1$ & 60.3$\pm3.6$ \\  
 \hline
 Specificity& 90.3$\pm0.3$ & 85.4$\pm4.0$ & 75.3$\pm2.3$ & 80.2$\pm2.3$ & 64.7$\pm6.5$ \\  
 \hline
    \end{tabular}
  \end{center}
\end{table*}

\section{Experimental Evaluation}
This section describes the experimental evaluation that includes data collection, implementation, training and results analysis.
\subsection{Data Collection and Annotation}
We recruit 5 participants (age range 23-35, average 27) during the month of Ramadan (between April 23rd, 2020 and May 23rd, 2020 ). The month of Ramadan is a holy month when religious Muslims honor their this special month by fasting from sunrise to sunset. During this time, fasting Muslims do not take any foods or drinks. We took advantages of this special occasion and recruited 5 religious Muslims who were fasting for a month. Upon the IRB exemption, the participants wore the Empatica E4 wristband from sunrise to sunset for a week-long study. Additionally, we collected one hour more data after the sunset which time participants were considered as very hydrated. During the data collection, participants were strictly prohibited to exercise, fast walk or doing physical labor that may cost extra water loss for them. Additionally, the device had to be taken off occasionally, both for showers (because it is not waterproof) and for charging (because there is not enough battery life for 72 hours of continuous running). We asked participants to keep a note regarding their thirst level during the day as follows: (i) after the last water intake (on sunrise) when is the first time he/she felt thirsty (transition from well-hydrated to hydrated), (ii) after the feeling thirsty first time, when he/she felt extremely thirsty as a second time (transition from hydrated to dehydrated), (iii) after feeling thirsty for second time, when he/she started feeling extremely thirst that he/she could not hold anymore and his/her tongue goes dry, (iv) when he/she started taking water/food by breaking the fasting (transition from very dehydrated to hydrated) and finally (v) when he/she finished taking food/water (transition from hydrated to very hydrated). Following the above instruction, we obtained proper annotation of 4 different hydration level of participants. Fig. \ref{fig:raw_signal} demonstrates the computation of indices of EDA in the time domain. This graph of data gathered over five minutes plots the subject's skin conductance level in microsiemens per centimeter square against time in seconds. The increase or decrease of skin surface's conductance in Fig. \ref{fig:raw_signal}  directly correlates to the levels of water in the subject's sweat glands.

\subsection{Machine Learning Models}
The Empatica device collects EDA readings at a rate of 4 Hz, or four samples per second. These samples are stored in a rolling window of eight samples. We consider a 5 seconds window as an activity window in which we first run two different deconvolution methods and then compute the 36 different features as described before.  As described before, we consider 4 different classes of hydration level: (1) Well Hydrated, (2) Hydrated, (3) Dehydrated, and (4) Very Dehydrated. Several machine learning models were trained in WEKA. Models produced include Random Forest classifiers, Decision Tree classifiers, Naive Bayes classifiers, BayesNet classifiers and Multilayer perceptron classifiers.

\begin{figure}
  \centering
  \includegraphics[width=\linewidth]{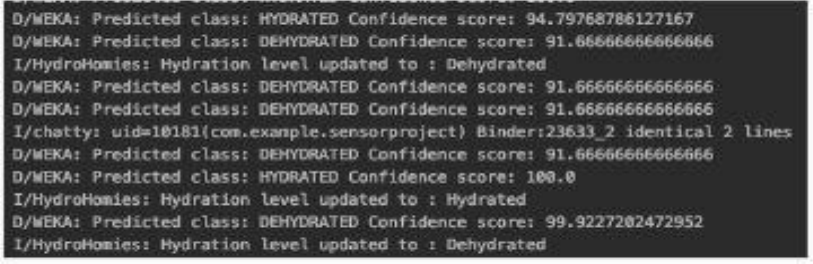}
  \caption{Decision Tree – Hydrated and Dehydrated}
  \label{fig:log}
\end{figure}
\subsection{Results Analysis}
After several trial and error, we achieved highest accuracy on the window size of 5 seconds and illustrated the comparisons of the results of different classifiers in Table \ref{tab:classification}. Table \ref{tab:classification} shows that 

The performance of these models with live data in the application was surprising when compared to their success ratings from training. The BayesNet classifier had the lowest confidence ratings – consistently as low as 60\% which made it the least reliable model. In the end, the Decision Tree model (with 84.5\%) accuracy was chosen as the final model for the project. It placed less stress on the computational load of the Android system and did not show signs of overfitting. Fig. \ref{fig:log} demonstrates some sample logs from running the application during testing. As can be seen, the Decision Tree model consistently has high confidence ratings.

\section{Conclusion}
To summarize, the Empatica E4 device shows promise in being used to detect hydration levels of subjects given the correct proportion of data varying over numerous subjects. The device has potential to be used in a completely non-invasive and easy-to-deploy way, unlike current methods of determining hydration level. The researchers have successfully implemented an application that reads data from an Empatica E4, uses it to calculate a subject's hydration level, and reports it to the subject in the application as well as through a notification. Our chosen machine learning model shows promise in being able to accurately classify subjects as hydrated or dehydrated, even with minimal data gathered only from one subject. The practical application of our research has certainly verified the promising future of wearable devices that have these multiple sensors and variables embedded within them to monitor real-time user activity.

\vspace{12pt}


\begin{thebibliography}{00}
\bibitem{1}	Survey of 3003 Americans, Nutritional Information Center, New York Hospital-Cornell Medical Center, April 14, 1998
 
\bibitem{2}	Ganio, M. S., Armstrong, L. E., Et. Al. “Mild dehydration impairs cognitive performance and mood of men,” British journal of Nutrition, 106(10), pp. 1535-1543, 2011.

\bibitem{3}	Armstrong, L. E., Et. Al. “Mild dehydration affects mood in healthy young women.” The Journal of nutrition, 142(2), 382-388, 2012.

\bibitem{4}	“Hydration Assessment of Athletes,” Oct. 2006. Accessed on: May 11,	2020. [Online].	Available: https://www.gssiweb.org/sports-science-exchange/article/sse-97-hydration-assessment-of-athletes.

\bibitem{5}	Mendelson, Y.; Dao, D.K.; Chon, K.H. Multi-channel pulse oximetry for wearable physiological monitoring. In Proceedings of the 2013 IEEE International Conference on Body Sensor Networks, Cambridge, MA, USA, 6–9 May 2013; pp. 1–6.

\bibitem{6}	H. F. Posada-Quintero, N. Reljin, A. Moutran, D. Georgopalis, E. C.-H. Lee, G. E. W. Giersch, D. J. Casa, and K. H. Chon, “Mild Dehydration Identification Using Machine Learning to Assess Autonomic Responses to Cognitive Stress,” Nutrients,vol. 12, no. 1, p. 42, 2019.

\bibitem{7}	F. Hernando-Gallego, D. Luengo, and A. Artes-Rodriguez, “Feature Extraction of Galvanic Skin Responses by Nonnegative Sparse Deconvolution,” IEEE Journal of Biomedical and Health Informatics,vol. 22, no. 5, pp. 1385–1394, 2018.

\bibitem{8}	Eibe Frank, Mark A. Hall, and Ian H. Witten (2016). The WEKA Workbench. Online Appendix for "Data Mining: Practical Machine Learning Tools and Techniques", Morgan Kaufmann, Fourth Edition, 2016.

\bibitem{9}Sollanek KJ, Kenefick RW, Cheuvront SN, Axtell RS. Potential impact of a 500-mL water bolus and body mass on plasma osmolality dilution. Eur J Appl Physiol 2011;111(9):1999.

\bibitem{10}Shanholtzer B, Patterson S. Use of bioelectrical impedance in hydration status assessment: reliability of a new tool in psychophysiology research. Int J Psychophysiol 2003 Sep;49(3):217-226.

\bibitem{11}Whitney EN, Rolfes SR. Understanding Nutrition. 9th ed. Belmont, CA: Wadsworth/Thomas Learning; 2002.

\bibitem{12} https://www.empatica.com/research/e4/

\bibitem{40}http://www.ledalab.de/

\bibitem{38}M. Benedek \& C. Kaernbach, A continuous measure of phasic electrodermal activity. Journal of Neuroscience Methods, 2010

\bibitem{39}M. Benedek \& C. Kaernbach, Decomposition of skin conductance data by means of nonnegative deconvolution, Psychophysiology, 2010.

\bibitem{alam1}Mohammad Arif Ul Alam, Nirmalya Roy, Single BSN-Based Multi-Label Activity Recognition, WristSense 2017.
\bibitem{alam2}Mohammad Arif Ul Alam, Nirmalya Roy, Archan Misra, Joseph Taylor, CACE: Exploiting Behavioral Interactions for Improved Activity Recognition in Multi-Inhabitant Smart Homes, 36th International Conference on Distributed Computing Systems, ICDCS 2016, Nara, Japan
\bibitem{alam3}Mohammad Arif Ul Alam, Nirmalya Roy, Sarah Holmes, Aryya Gangopadhyay, Elizabeth Galik, Automated Functional and Behavioral Health Assessment of Older Adults with Dementia, IEEE Conference on Connected Health: Applications, Systems and Engineering Technologies, CHASE 2016, Washington DC, USA

\end{thebibliography}
\end{document}